# Interplay between the range of attractive potential and metastability in gas-liquid nucleation

Rakesh S. Singh, Mantu Santra and Biman Bagchi[*]

Solid State and Structural Chemistry Unit,

Indian Institute of Science, Bangalore 560012, India.

[*]Email: bbagchi@sscu.iisc.ernet.in

## *Abstract*

We find an interesting interplay between the range of the attractive part of the interaction potential and the extent of metastability (as measured by supersaturation) in gas-liquid nucleation. We explore and exploit this interplay to obtain new insight into nucleation phenomena. Just like its dependence on supersaturation ($S$), the free energy barrier of nucleation is found to depend strongly on the range of the interaction potential. Actually, the entire free energy surface, $F(n)$, where $n$ is the size of the liquid-like cluster, shows this dependence. The evidences and the reasons for this strong dependence are as follows. (i) The surface tension increases dramatically on increasing the range of interaction potential. In three dimensional Lennard-Jones system, the value of the surface tension increases from 0.494 for a cut-off of 2.5 σ to 1.09 when the full range of the potential is involved. In two dimensional LJ system, the value of the line tension increases from 0.05 to 0.18, under the same variation of the potential range. (ii) The density of the gas phase at coexistence decreases while that of the liquid phase increases substantially on increasing the range of the interaction potential. (iii) As a result of the above, at a given supersaturation $S$, the size of the critical nucleus and the free energy barrier both increase with increase in the range of interaction potential. (iv) Surprisingly, however, we find that the functional form predicted by the classical nucleation theory (CNT) for the dependence of the free energy barrier on the size of the nucleus to remain valid except at the largest value of $S$ studied. (v) The agreement between CNT prediction and simulated values of the barrier is supersaturation dependent and worsens with increase in the range of interaction potential, and increases above 10 $k_BT$ at the largest supersaturation that could be studied.

## I. Introduction

For over half-a-century, the study of nucleation and growth has remained a subject of great interest among workers from different branches of science and among various kinds of nucleation in different systems, the one from gas to liquid has been perceived as a bench mark example for other studies. Nucleation of the liquid phase from a supersaturated vapor phase is usually an activated process that involves the formation of a thermodynamically stable liquid-like critical nucleus within the metastable parent vapor phase. The classical nucleation theory (CNT) **[1-3]** provides a simple description for homogeneous nucleation. It assumes that a small droplet of the new stable phase first forms as a result of spontaneous density fluctuation. This small droplet or embryo then grows in a sea of parent metastable bulk phase by addition of single molecules.

Many statistical mechanical calculations and simulations are carried out in grand canonical ensemble (at fixed chemical potential $\mu$, volume $V$ and temperature $T$). The grand canonical potential for a coexisting two phase system consisting of a spherical liquid droplet of radius $R$ in a sea of metastable vapor phase is

$$\Omega = -P_l V_l - P_v V_v + 4\pi R^2 \gamma_\infty, \tag{1}$$

where $V_l$ and $V_v$ are the volumes of the liquid droplet and the vapor. $\gamma_\infty$ is the liquid-vapor interfacial surface tension, assumed to be given by that of the planar interface at coexistence at the same temperature at which nucleation is taking place.

We then can write the grand canonical potential for formation of a spherical liquid droplet of radius $R$ as

$$\Delta\Omega(R) = -\frac{4\pi}{3}R^3\Delta P + 4\pi R^2\gamma_\infty, \tag{2}$$

where $\Delta P = P_l - P_v$. $P_l$ is the pressure of the bulk liquid and $P_v$ is the pressure of the bulk vapor. By imposing the stationary condition for the maximum, $\partial\Delta\Omega(R)/\partial R = 0$, we get the expression for radius of the critical droplet and barrier height for nucleation

$$R^* = \frac{2\gamma_\infty}{\Delta P} \quad \text{and} \tag{3}$$

$$\Delta\Omega^* = \frac{16\pi}{3}\frac{\gamma_\infty^3}{(\Delta P)^2}. \tag{4}$$

The last expression predicts a stronger dependence of the free energy barrier on the surface tension between the two phases than on the free energy difference between the two phases. This difference in relative dependencies plays important role at large supersaturation, as the surface tension is often predicted to decrease with supersaturation **[4]**. In two dimensions, one essentially finds similar expressions for the critical radius and the free energy barrier.

The above expressions are quite general and have been applied to a variety of first order phase transitions. For gas-liquid nucleation, if we further assume that the liquid is incompressible and the vapor ideal, the pressure difference can be given as $\Delta P = \rho_l k_B T \ln(P/P_{coex}) = \rho_l k_B T \ln S$, where $S = P/P_{coex}$ is the supersaturation in terms of the pressure of the system (described below). Additionally, in computer simulation studies it is convenient to define the size of a cluster in terms of number of particles $n$ that it contains. Under these conditions, the expressions for the critical number and the free energy barrier are given by the following equations

$$n^* = \frac{32\pi\gamma_\infty^3}{3\rho_l^2 (k_B T \ln S)^3} \:,  \tag{5}$$

$$\Delta\Omega^* = \frac{16\pi\gamma_\infty^3}{3\rho_l^2 (k_B T \ln S)^2} \:.  \tag{6}$$

Here $\rho_l$ is the density of the bulk liquid at coexistence. The quantity $S$ gives a quantitative measure of the metastability of vapor phase. As mentioned earlier, it is defined as $S = P/P_{coex}$, where $P_{coex}$ is the equilibrium pressure at coexistence, and $P$ is the pressure of the system at which nucleation occurs. In grand canonical; ensemble it is more convenient to follow the chemical potential definition of supersaturation, defined as $S = \exp(\beta\Delta\mu)$, where $\Delta\mu = \mu - \mu_{coex}$, with $\mu$ the imposed chemical potential and $\mu_{coex}$ is the chemical potential at coexistence.

The validity of CNT expression (as developed by Becker, Döring and Zeldovich) of the free energy barrier for gas-liquid nucleation has been widely tested by carrying out both detailed computer simulation and experimental studies **[5-8]**. In computer simulation studies of gas-liquid nucleation, it has been found that for three dimensional (3D) systems, the CNT overestimates the actual free energy barrier by about 4-5 $k_B T$. Recent computer simulation studies of gas-liquid nucleation in two dimension (2D) by Santra *et al.* **[5]** have shown that CNT *underestimates* the barrier by as much as 70% (at supersaturation $S$ = 1.1 and reduced temperature $T^*$ = 0.427 ). The size of the critical cluster is also vastly underestimated by CNT in 2D gas-liquid nucleation. The latter authors have shown that the extent of agreement between simulation and CNT prediction depends strongly on the range of interaction potential (usually truncated at $2.5\sigma$) used in simulation. Interestingly, the above mentioned study observed that, in the case of 2D gas-liquid nucleation, as the range of the interaction potential increases the agreement between simulation and CNT predictions of free energy barrier actually *improves*. The shape corrections also have a remarkable effect on the improvement of the disagreement between simulation and CNT prediction --- the shape becomes more circular as the range of interaction potential increased. The strong dependence of line tension on the range of interaction potential has also been observed. The assumption of the circular shape of the critical cluster was found to be inadequate but only at large supersaturation.

Recent experiments on gas-liquid nucleation find a much higher nucleation rate than what predicted by CNT **[6, 7]**. Strey and co-workers have studied nucleation of argon in supersonic nozzle at large supersaturation (the ratio of the actual vapor pressure and the vapor pressure at coexistence at the same temperature), $S \geq 40$, and found a rate equal to $10^{17} \, cm^{-3} s^{-1}$. The CNT prediction is 11-13 orders of magnitude lower **[6]**. An earlier study on the nucleation rate of argon in cloud chambers at lower supersaturation $S = 10$ and $T = 52\text{-}59$ K and reported nucleation rate equal to $10^{7} \, cm^{-3} s^{-1}$, where as CNT predicts a rate of $10^{-13} \, cm^{-3} s^{-1}$ **[7]**. They have argued that classical nucleation theory fails because it overestimates both the critical cluster size and the excess internal energy of the critical cluster. Also, note that there is no truncation of potential in experimental systems.

One of the possible reasons for the observed lack of validity of CNT is that it assumes the critical nucleus as (i) spherical and (ii) bulk liquid like and (iii) uses the surface tension from the free energy cost of the formation of a planar interface between the stable bulk liquid and vapor phases. However, both in computer simulations and in density functional theoretical studies one finds that due to finite size of the critical cluster, the density of the critical cluster at the core is less than the bulk liquid phase and the interface is diffuse **[9-13]**. That is, the critical cluster is dressed by an intermediate density phase (neither vapor- nor liquid-like) which reduces the surface tension. Thus, CNT overestimates both the surface tension, $\gamma$ and $\Delta P$. We have discussed already that the shape of the critical cluster plays an important role in 2D nucleation. In 3D one can expect that shape correction will not be as important as in 2D due to larger surface tension in 3D which does not allow large fluctuations in shape of the critical cluster.

In order to check the generality of theoretical predictions and the cause of discrepancy between theoretical and experimental/simulation studies, we have studied the effects of the range of interaction potential on the nature of the liquid-vapor interface, the surface tension, the free energy barrier for nucleation, and the size of the critical nucleus in gas-liquid nucleation. Although there have been several studies on the dependence of the gas-liquid surface tension on the range of interaction potential **[14,15]**, a systematic study, however, of dependence of 3D gas-liquid nucleation scenario on range of interaction is still lacking. In crystallization the strong effect of range of interaction potential has already been found by molecular dynamics simulations **[16]**. The crystal structure changes with change in the range of the interaction potential. For short range interaction potential *fcc* crystals form and for long range interaction potential *bcc* crystals form. In the case of gas-liquid nucleation, the liquid phase becomes unstable if the range of the attractive potential is too small. The extreme example is a square well potential with very narrow attractive window. In this case, the gas-liquid phase transition may not exist. Thus, if we decrease the range of the attractive part of the potential, we in effect move the system closer to its gas-liquid critical point. So, we expect the gas-liquid surface tension to decrease and the liquid to have lower density at a given temperature. However, somewhat surprisingly, no detailed study of this effect has been studied or reported.

The motivation of the study partly originates from the observation that one can vary the extent of accessible values of supersaturation $S$ by varying the range of the attractive part of the potential. This then provides us with a wider range of parameters to test the CNT and understand its limitations. For example, when the range of interaction potential is small (that is, the potential is narrow), the maximum supersaturation that can be accessed in simulations of LJ spheres is ~ 2.5 (a value well-documented in literature). However, with the full range included, the maximum $S$ is now close to 10. Now, one can compare the nucleation scenarios in both the cases and arrive at certain important conclusions. For example, we find, somewhat surprisingly, that the functional form predicted by the classical nucleation theory (CNT) for the dependence of the free barrier energy on the size of the nucleus remains valid till the kinetic spinodal limit (the value of $S$ where the free energy of the largest cluster becomes zero). However, the agreement between CNT prediction and simulated values of the barrier worsens with the increase of $r_{cut}$, and the barrier increases to above 10 $k_BT$ at the largest supersaturation that could be studied.

The organization of the rest of the paper is as follows. In the next section (section II), we present the model system and the simulation used in this work. In Section III contains a discussion and also a comparison with 2D nucleation. Section IV contains concluding remarks.

## II.     Model system and computational details

The model system studied consists of particles interacting with each other with Lennard-Jones (LJ) potential

$$v_{LJ} = 4\varepsilon\left[\left(\frac{\sigma}{r}\right)^{12} - \left(\frac{\sigma}{r}\right)^{6}\right]. \tag{7}$$

We have studied systems with above potential truncated and shifted at different radii $(r_{cut} = 2.5\sigma, 3.5\sigma, 5\sigma \text{ and } 7\sigma)$. The truncated and shifted potential is given by

$$\begin{aligned} v(r) &= v_{LJ}(r) - v_{LJ}(r_{cut}) & r \leq r_{cut} \\ &= 0 & r > r_{cut} \end{aligned}, \tag{8}$$

where $r_{cut}$ is the cutoff radius. All our studies have been carried out at reduced temperature $T^* = 0.741$. Following the definition by Stillinger [17] and the successful modification by Frenkel and co-worker [18], we consider a particle to be liquid-like if it has more than four nearest neighbors within a cutoff distance of $1.5\sigma$. Liquid-like particles that are connected by neighborhood (within the cutoff distance $1.5\sigma$) form liquid-like clusters. We have used Transition Matrix Monte Carlo (TMMC) method [19], a very efficient non-Boltzmann sampling scheme, to obtain the free energy surfaces. For free energy calculation the simulations are performed in $\mu VT$ ensemble. In this work, the supersaturation (S) is defined as $S = \exp(\beta\Delta\mu)$, where $\Delta\mu = \mu - \mu_{coex}$, with $\mu$ the imposed chemical potential and $\mu_{coex}$ is the chemical potential at coexistence.

To obtain the activity coefficient at coexistence for different range of interaction potential we have performed Grand Canonical Transition Matrix Monte Carlo (GC-TMMC) simulation studies in a cubic box near a trial coexistence activity. Resulting probability density as a function of density of the system was then extrapolated with respect to $\xi^*$ to obtain the exact coexistence activity [19, 20]

$$\ln \prod(\rho,\mu') = \ln \prod(\rho,\mu) + \beta\rho V(\mu'-\mu). \tag{9}$$

To compute interfacial vapor-liquid surface tension at coexistence the coexistence chemical potential needs to be known. Using GC-TMMC method first we have calculated the coexistence chemical potential for different range of interaction potential and then at those coexistence points the free energy of formation of a vapor-liquid interface has been computed [21]. From these results, the interface free energy per unit area, $\gamma^*$, has been obtained. The reduced surface tension is defined as $\gamma^* = \gamma\sigma^2/\varepsilon$.

## III. Results and discussion

In this section we discuss main results of the paper and also the comparison with the 2D gas-liquid nucleation.

### A. Dependence of gas-liquid coexistence interface and surface tension on the range of interaction potential

Surface tension is the one of the most fundamental properties of the interface and is required for the quantitative prediction of both the nucleation barrier as well as the size of the critical nucleus. The dependence of surface tension on range of interaction potential has been discussed extensively in the literature [14, 15]. Here we shall provide some more insight in this dependence and show how it scales with the range of interaction potential.

First, we have computed the density profile at liquid-vapor coexistence for different range of interaction potentials. **Fig. 1** shows the gradient of the density profile along the planar interface (the interface has been created along Z axis). We see from the figure that *the height of the density gradient increases with increase of the range of interaction potential*. Interestingly, the *width of the interface remains almost constant*. The density of the liquid phase at coexistence increases moderately while that of vapor phase decreases substantially on increasing the range of the interaction potential (see **Table I**). That is, on increasing the range of interaction potential, the difference between the densities of bulk liquid and vapor phases increases and which makes the interface sharper. This also gives rise to a sharp rise in the value of the surface tension, as discussed below.

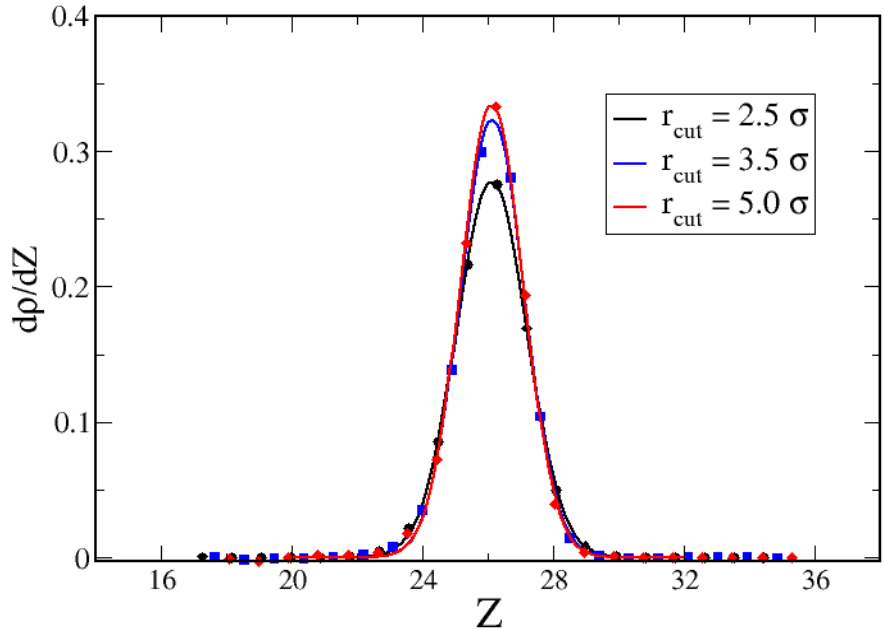

**Figure – 1. Plot of derivative of density profile with respect to box length along Z-axis at different cutoff. The dots indicate the simulation data and continuous lines indicate the Gaussian fit of the simulation data points.**

In **Table 1** the bulk densities of liquid and vapor phases, coexistence pressure, activity coefficient at coexistence and surface tension for different cut-off ranges, $r_{cut}$, are reported. The value of the surface tension increases becomes almost double on increasing cut-off from 2.5 $\sigma$ to 5.0 $\sigma$. On the other hand, two dimensional LJ system shows even stronger dependence of the line tension on the range of interaction potential (it increases from 0.05 to 0.18 on increasing the cut-off from 2.5 $\sigma$ to 7.0 $\sigma$ at $T^*=0.427$).

**TABLE I. The, coexistence bulk densities of vapor $\left(\rho_v^*\right)$, liquid $\left(\rho_l^*\right)$, coexistence pressure ($P_{coex}$), coexistence activity $\left(\xi_{coex}^*\right)$ and vapor-liquid surface tension $\left(\gamma^*\right)$ for 3D LJ system for different range of interaction potentials ($r_{cut}$).**

| $r_{cut}$ | $\rho_v^*$ | $\rho_l^*$ | $P_{coex}$ | $\xi_{coex}^*$ | $\gamma^*$ |
|---|---|---|---|---|---|
| 2.5 | 0.0115 | 0.766 | 0.00783 | 0.00973 | 0.494 |
| 3.0 | 0.00678 | 0.792 | 0.00475 | 0.00608 | 0.672 |
| 3.5 | 0.00517 | 0.804 | 0.00366 | 0.00474 | 0.783 |
| 4.0 | 0.00444 | 0.806 | 0.00317 | 0.00412 | 0.857 |
| 5.0 | 0.00383 | 0.818 | 0.00274 | 0.00358 | 0.945 |

The dependence of surface tension ($\gamma^*$) on the cut-off range, $r_{cut}$, is shown in **Fig. 2(a)**. Previous theoretical study by Korochkova *et al*. **[15]** suggests that the dependence of surface tension on $r_{cut}$ scales as $\Delta\rho^2/r_{cut}^2$ ($\Delta\rho = \rho_l^* - \rho_v^*$), however, surprisingly, we find that our simulation data scales as $1/r_{cut}^2$. That is, there is no explicit $\Delta\rho$ dependence in the scaling relation. In figure, the black circles show the simulation data and red line indicates the fit of the simulation data by the function

$$\gamma(r_{cut}) = \gamma(\infty) - \frac{B}{r_{cut}^2}, \tag{10}$$

where $\gamma(\infty) = 1.092$ and $B = 3.756$ at $T = 0.741$. The scaling relation is plotted in **Fig. 2(b)**. Extrapolation of the surface tension data up to $r_{cut} \to \infty$ indicates that for full LJ potential the surface tension would be 1.092, which is in excellent agreement with the extrapolated values of the previous work **[14]**.

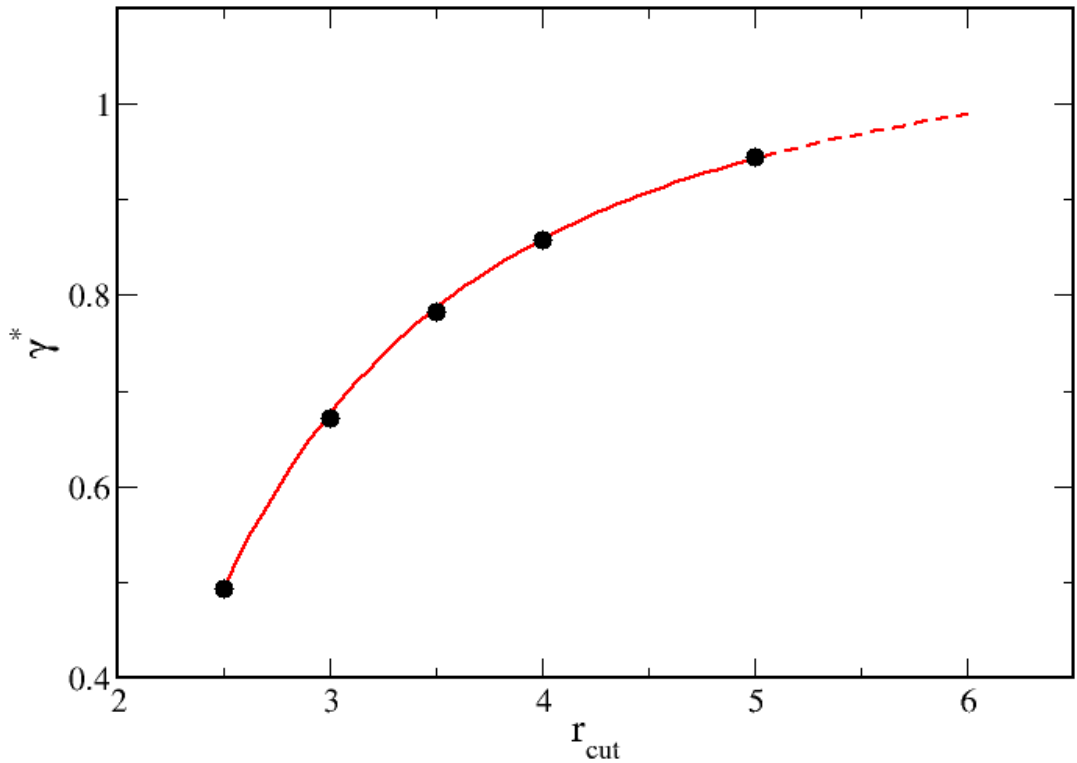

**Figure – 2 (a). Dependence of surface tension on the range of interaction potential ($r_{cut}$). The black circles indicate the simulation data and the continuous red line shows the fit by Eq. (10).**

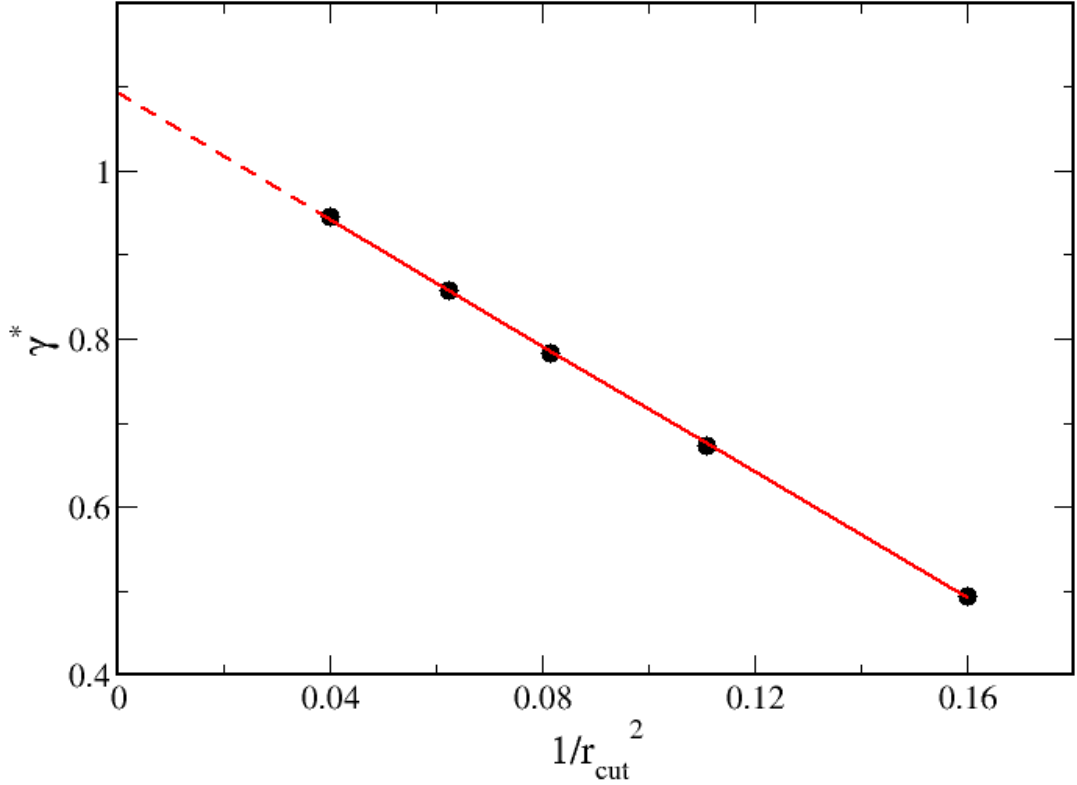

**Figure – 2(b). Dependence of surface tension on ($1/r_{cut}^2$). The red line indicates the linear fit of simulation data with slope -3.7561 and intercept 1.0923.**

As shown in **Fig. 2(a)** and **Fig. 2(b)**, the system exhibits a strong dependence of surface tension on the cut-off range $r_{cut}$. As already mentioned, this strong dependency arises due to increase in the difference of the liquid and vapor densities on increasing $r_{cut}$. As surface tension strongly depends on the gradient of the density profile of the interface and it drastically increases on increasing $r_{cut}$. From the surface tension values at different $r_{cut}$ and scaling plot, we can see that the increase of surface tension with $r_{cut}$ slows down and eventually converges at larger value of $r_{cut}$. The dependency of surface tension on supersaturation is discussed in **section III C.**

## B. Dependence of the free energy barrier and the critical nucleus on the range of interaction potential and comparison with CNT

The computed free energy barriers for gas-liquid nucleation as a function of $S$ for three different ranges of interaction potential $(r_{cut} = 2.5\sigma, 3.5\sigma$ and $5\sigma)$ are shown in **Fig. 3 (a).**

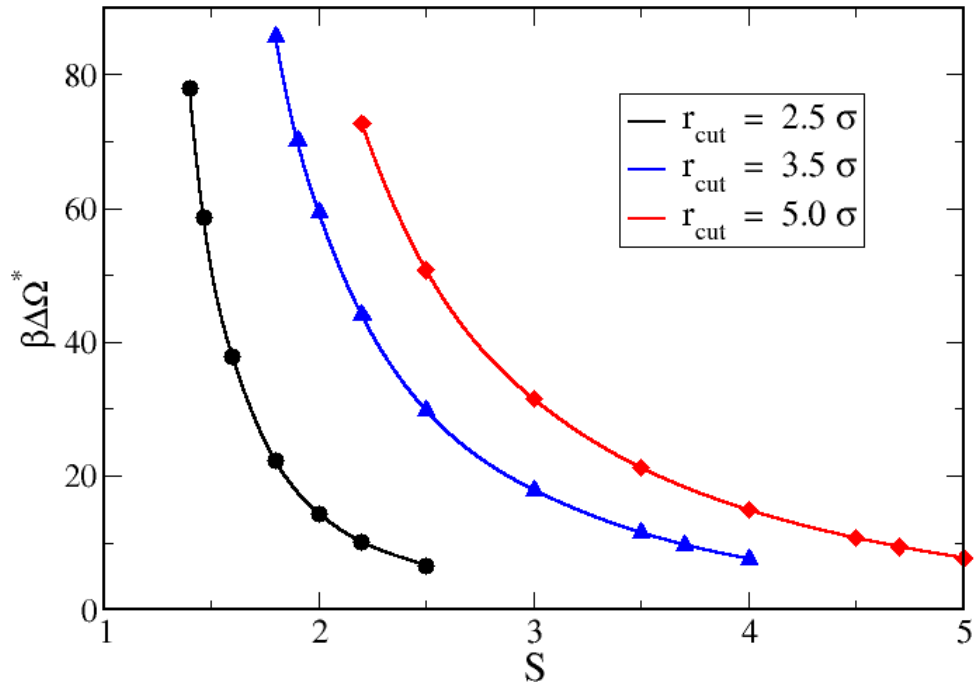

**Figure – 3 (a). Dependence of the free energy barrier for gas-liquid nucleation on supersaturation ($S$) at three different values of range of interaction potential ($r_{cut}$), 2.5, 3.5 and 5.0$\sigma$, respectively.**

The nucleation barrier shows the strong dependence on the range of interaction potential. Actually, the entire free energy surface, $F(n)$, where $n$ is the size of the liquid like cluster shows the strong dependence on $r_{cut}$. As expected, at low supersaturation the barrier decreases rapidly with S, whereas at high supersaturation the rate of change becomes slower as expected.

The reason behind the strong $r_{cut}$ dependence is that as we increase the range of the interaction potential the surface tension increases dramatically, as discussed in **section III A**. In 3D Lennard-Jones system, the value of the surface tension increases from 0.494 for a cut-off of 2.5 σ to 0.945 for cut-off of 5.0 σ. As a result, at a given supersaturation the free energy barrier increases drastically with increase in the range of interaction potential.

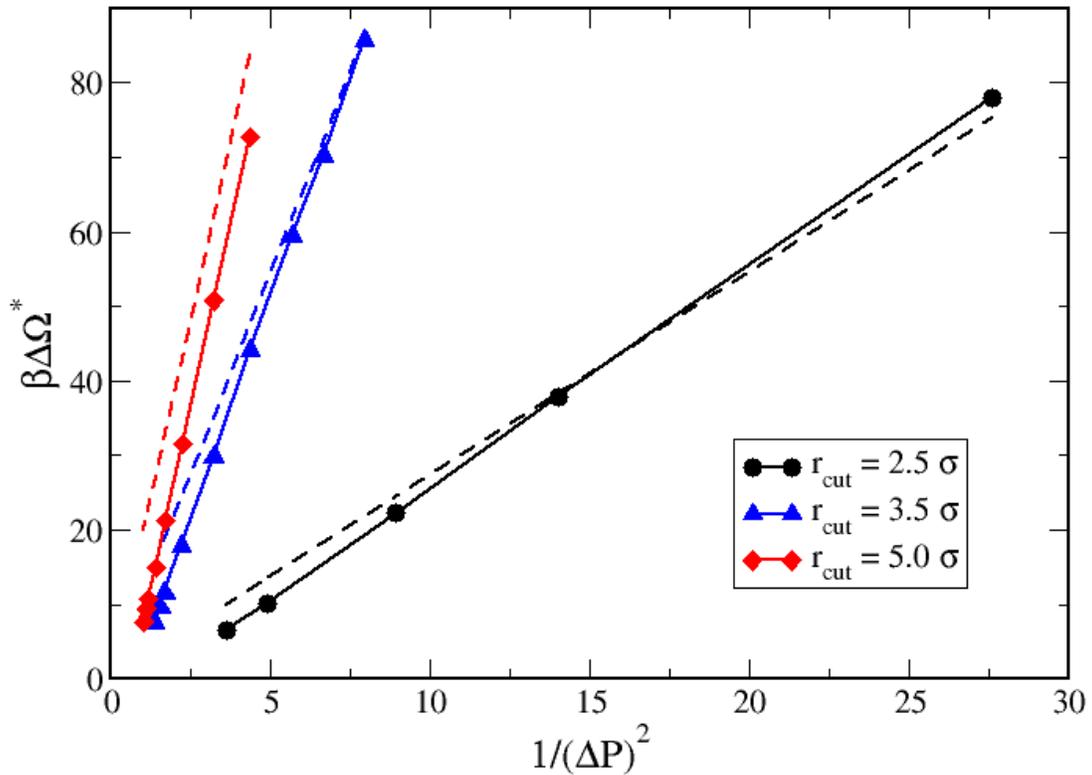

**Figure – 3 (b). Dependence of the free energy barrier for nucleation on $1/(\Delta P)^2$ at three different values of range of interaction potential ($r_{cut}$), 2.5, 3.5 and 5.0 $\sigma$, respectively. Dotted lines indicate the CNT predictions. The corresponding colors for different cut-offs are indicated in the inset.**

The conventional choice to show the dependency of the nucleation barrier on supersaturation is to plot the free energy barrier against $1/(\Delta P)^2$, as according to CNT the barrier height is linear function of this quantity. The pressure difference is computed using following equation,

$$\Delta P = k_B T \int_{\xi_{coex}}^{\xi} \left[ \frac{\rho_l(\xi')}{\xi'} - \frac{\rho_v(\xi')}{\xi'} \right] d\xi' \qquad (11)$$

Here, $\xi_{coex}$ is the activity coefficient at coexistence and $\rho_v$ and $\rho_l$ are the density of bulk vapor and liquid phase, respectively. We have numerically integrated Eq. (11) after obtaining the bulk densities from simulation at different activity coefficients.

In **Fig. 3 (b)** the CNT predictions are indicated by dotted lines. We find that the functional form predicted by the classical nucleation theory (CNT) for the dependence of the free barrier energy on the supersaturation remains valid even at very large *S* studied. Literature studies show that the CNT overestimates the nucleation barrier by a constant value which is independent of supersaturation but depends upon temperature **[22]**. Interestingly, here we find that there is a crossover of the free energy predicted by CNT and simulation. At high supersaturation CNT overestimates the barrier height as found in previous studies, but it underestimates at low supersaturation. The density functional theory (DFT) study by Oxtoby *et al.* **[23]** shows a similar crossover. This may be due to the fact that the effective surface tension for large droplets is greater than that of the planar interface, but it becomes smaller for small droplets. Also in experiments, a crossover in rate between CNT and experimental data has been observed on going from small to large super-cooling.

As the range of interaction potential increases the discrepancy between simulation and CNT predictions of the free energy barrier increases. The discrepancy depends upon many factors. One of them is the region of supersaturation we studied. Here we should mention that the regions of supersaturation for different cut-off are different. **Fig. 4** depicts the dependence of free energy barrier on $r_{cut}$ at a fixed supersaturation, $S = 2.2$. The free energy barrier increases from 10 to 73 $k_B T$ on changing the range of interaction potential from $r_{cut} = 2.5\,\sigma$ to $5.0\,\sigma$. The discrepancy in barrier with CNT prediction at $r_{cut} = 2.5\sigma$ is approximately $3.5\,k_B T$ and at $r_{cut} = 5.0\sigma$ is approximately $10.0\ k_B T$. This is opposite to the case of 2D nucleation, where the discrepancy between the simulation and CNT prediction decreases with increase in the range of interaction potential **[5]**.

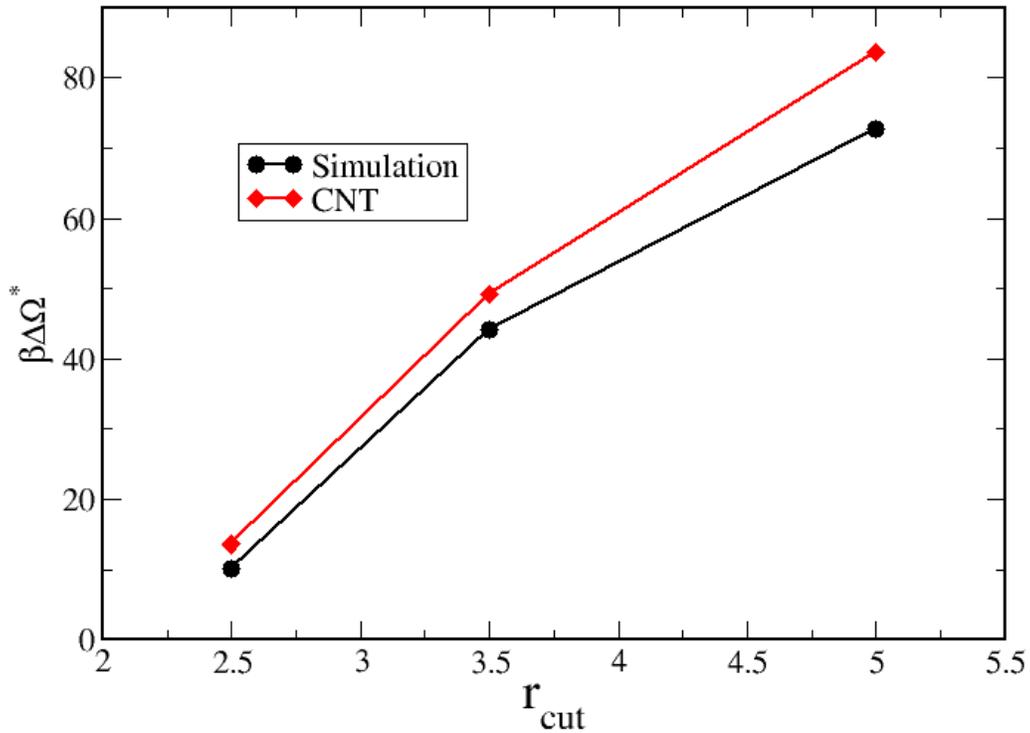

**Figure – 4. Change of the free energy barrier for nucleation with range of interaction potential ($r_{cut}$) at a fixed supersaturation, $S = 2.2$.**

The dependence of the size of the critical nuclei on supersaturation ($S$) is shown in **Fig. 5 (a).** Following the convention, we have plotted the critical nucleus size versus $1/(\Delta P)^3$ (as in CNT for incompressible liquid the barrier height is linear function of this quantity) in **Fig. 5 (b)**. The figure shows the strong dependence of the critical cluster size on the range of interaction potential. In **Fig. 6** the dependence of critical nucleus on $r_c$ at a fixed supersaturation, $S = 2.2$ is shown. At a given supersaturation $S = 2.2$, the size of the critical nucleus increases with increase in the range of interaction potential. The size of the critical nucleus increases from 20 at $r_{cut} = 2.5\ \sigma$ to 182 at $r_{cut} = 5.0\ \sigma$. As we know that the size of the critical cluster strongly depends on surface tension, we attribute again this strong dependence to the increase in the surface tension on increasing the range of the interaction potential.

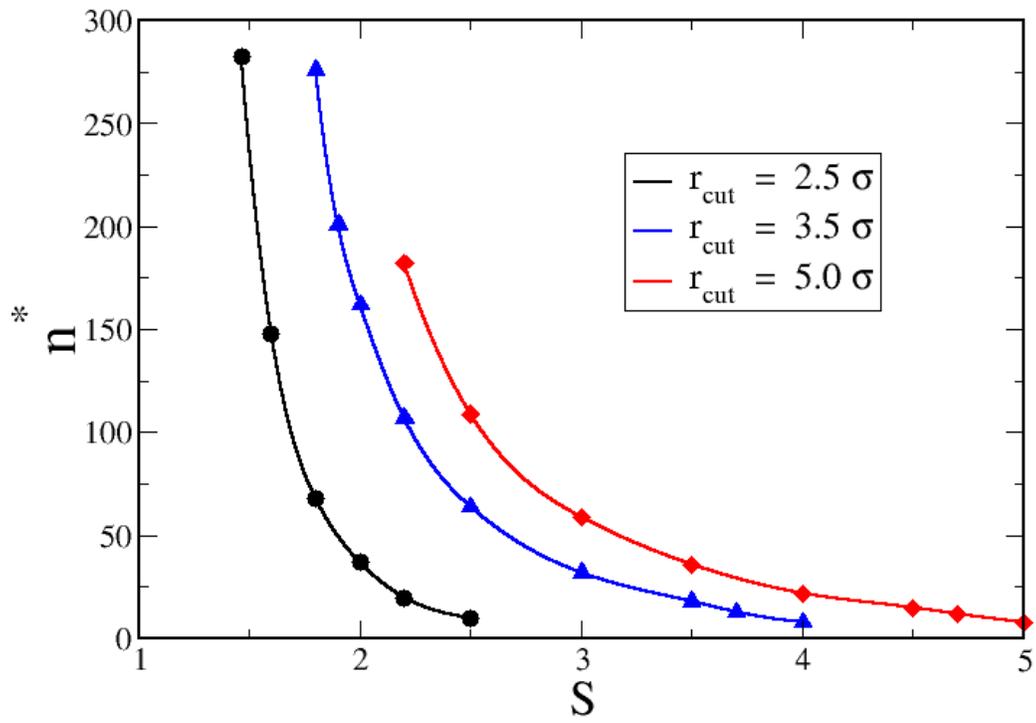

**Figure – 5(a).** Dependence of the size of the critical nuclei on supersaturation (*S*) for $r_{cut} = 2.5\sigma$, $3.5\sigma$ and $5.0\sigma$, respectively.

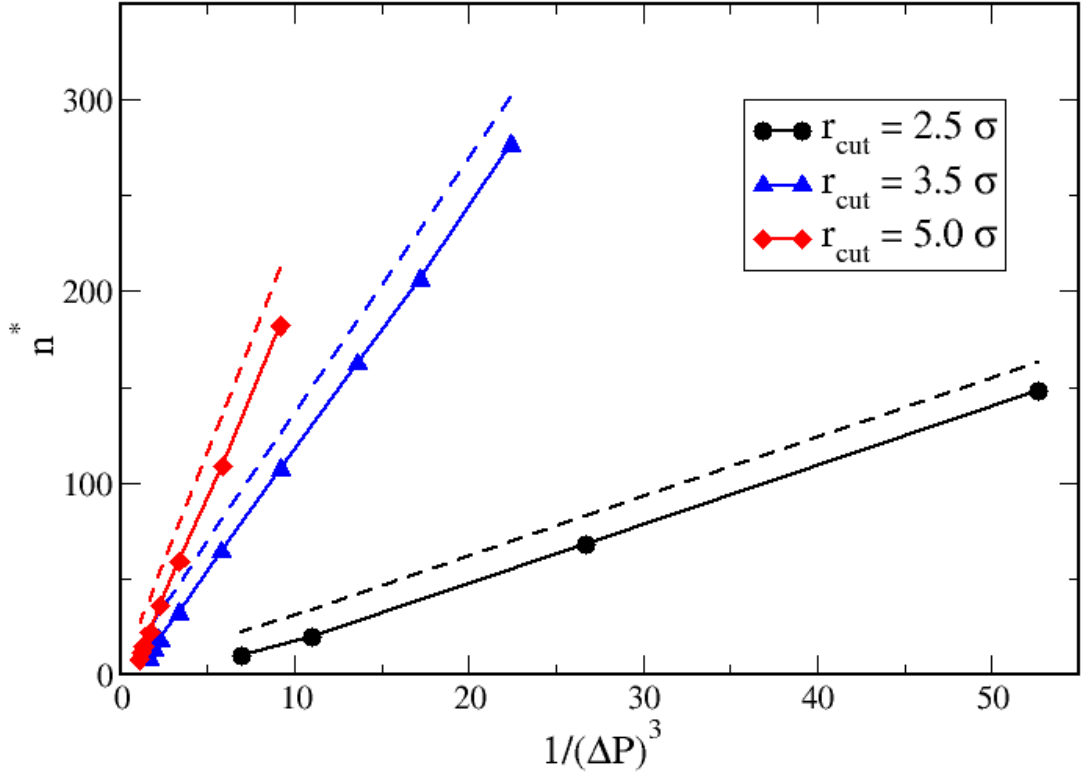

**Figure – 5 (b).** Dependence of the critical nucleus size on $1/(\Delta P)^3$ at three different values of range of interaction potential ($r_{cut}$), 2.5, 3.5 and 5.0 $\sigma$, respectively. Dotted lines indicate the CNT predictions.

As shown in **Fig. 5 (b)**, the functional form of CNT expression for the dependence of critical cluster size on supersaturation remains valid even at large supersaturation. However, there is almost a constant offset between simulation and CNT prediction of the size of critical cluster. CNT always overestimates the size of the critical cluster. We must point out that, while, the size of the critical cluster is highly sensitive to the definition of liquid-like particles the main results and the conclusions should remain valid with an alternative definition of order parameter.

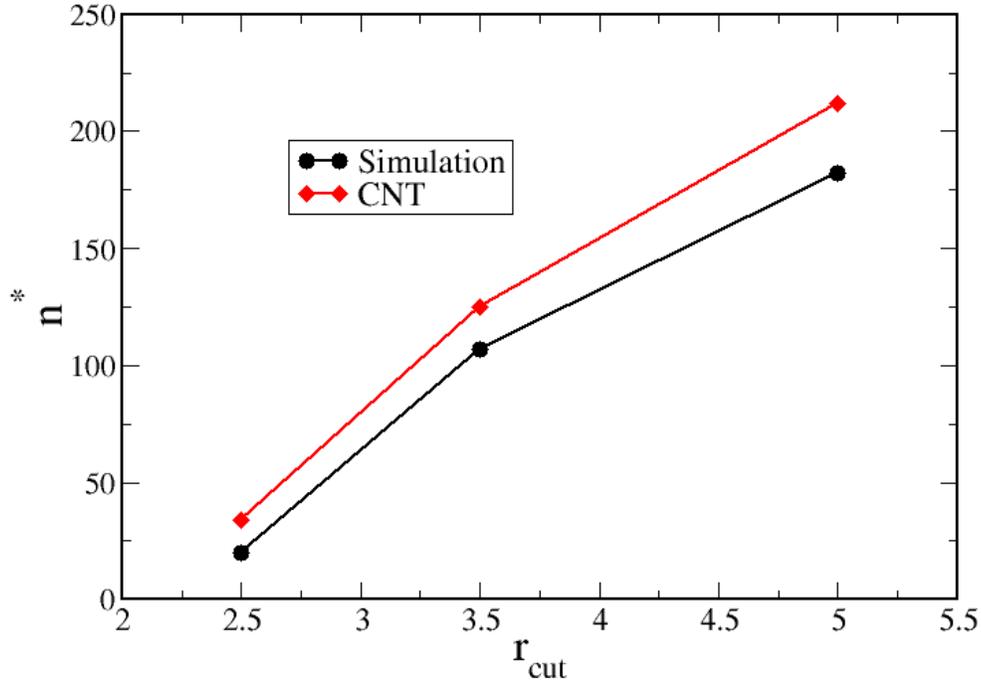

**Figure – 6.** Change of the critical nuclei with range of interaction potential ($r_{cut}$) at a fixed supersaturation, $S = 2.2$.

## IV. Conclusion

Let us first summarize the main results of the paper. In order to understand the mechanism of nucleation at large supersaturation, particularly at the limit of metastability near the spinodal point, we investigated various aspects of the gas-liquid nucleation phenomena in a simple model system consisting of spheres interacting with each other via the Lennard-Jones intermolecular potential. Some of the results obtained are quite surprising and we summarize them below.

We find an interesting interplay between the range of interaction potential and the extent of metastability (as measured by supersaturation). The free energy barrier of nucleation is found to depend strongly on the range of the interaction potential. The surface tension increases dramatically on increasing the range of interaction potential. In three dimensional Lennard-Jones system, the value of the surface tension increases from 0.494 for a cut-off of 2.5 σ to 1.09 for full range of interaction potential. In two dimensional LJ system, the value of the line tension increases from 0.05 to 0.18, under the change of the potential range from 2.5 σ to full range. The density of the gas phase at equilibrium coexistence decreases while that of the liquid phase increases substantially on increasing the range of the interaction potential. However, the width of the interface remains approximately the same. This increases the surface tension dramatically. As a result, at a given supersaturation *S,* the size of the critical nucleus and the free energy barrier both increase with increase in the range of interaction potential. Interestingly, this makes the premises of CNT more applicable. The functional form predicted by the classical nucleation theory (CNT) for the dependence of the free energy barrier on the supersaturation is found to remain valid except at the largest value of *S* studied. In terms of absolute values of numbers, the agreement between CNT prediction and simulated values of the barrier worsens with the increase in cut-off radius $r_{cut}$. This discrepancy increases to above 10 $k_BT$ at the largest supersaturation studied. The reason for such failure is not understood quantitatively. It is possible that $\Delta P$ fails to represent the pressure difference of the droplet as the liquid nucleus is likely to become less compact and fully liquid-like at large metastability. We found earlier that the coordination number of atoms of the liquid embryo decreases with metastability. While this can explain the trend qualitatively, we have not yet been able to quantify the dependence. This requires further work.

## Acknowledgement

We thank DST, India for partial support of this work. B.B. thanks DST for a JC Bose Fellowship.